# Blocking of conducting channels widens window for ferroelectric resistive switching in interface-engineered $Hf_{0.5}Zr_{0.5}O_2$ tunnel devices

Milena Cervo Sulzbach, Saúl Estandía, Jaume Gàzquez, Florencio Sánchez, Ignasi Fina*, Josep Fontcuberta*

Institut de Ciència de Materials de Barcelona (ICMAB-CSIC), Campus UAB, Bellaterra, Catalonia 08193, Spain

E-mail: ifina@icmab.es, fontcuberta@icmab.cat

Films of $Hf_{0.5}Z_{0.5}O_2$ (HZO) contain a network of grain boundaries. In (111) HZO epitaxial films on (001) $SrTiO_3$, for instance, twinned orthorhombic (o-HZO) ferroelectric crystallites coexist with grain boundaries between o-HZO and a residual paraelectric monoclinic (m-HZO) phase. These grain boundaries contribute to the resistive switching response in addition to the genuine ferroelectric polarization switching and have detrimental effects on device performance. Here, it is shown that, by using suitable nanometric capping layers deposited on HZO film, a radical improvement of the operation window of the tunnel device can be achieved. Crystalline $SrTiO_3$ and amorphous $AlO_x$ are explored as capping layers. It is observed that these layers conformally coat the HZO surface and allow to increase the yield and homogeneity of functioning ferroelectric junctions while strengthening endurance. Data show that the capping layers block ionic-like transport channels across grain boundaries. It is suggested that they act as oxygen suppliers to the oxygen-getters grain boundaries in HZO. In this scenario it could be envisaged that these and other oxides could also be explored and tested for fully compatible CMOS technologies.

**Introduction**

Data storage and computing have been pivotal in humankind evolution and are ubiquitous in the Internet of Things (IoT) society. Nowadays, fast-access data are stored in charge-based elements with most advanced devices having node sizes around roughly 18 nm to 15 nm. However, this technology is no longer shrinking the node size at the same pace as in previous decades. Moreover, the technology, such as 5G and machine-learning, onslaught for faster and even more data intensive requirements and urges for new memory types. Dynamic random access memory (DRAM) is a key building block in such systems, it is fast and cheap, however it also presents important drawbacks.[1] DRAM is a volatile memory, meaning that data are lost when the power is shut off. On the other hand, flash (NAND flash) memory is non-volatile and therefore retains data when power is off. However, in spite of its popularity, flash cannot be the ultimate technology for data storage and computing due to speed, endurance and power limitations.[2] DRAM and NAND flash store bits of information in the form of electric charge. Thus, as their size shrinks, the amount of stored charge reduces (and leaks), compromising its retention and reading and, ultimately, its function. Resistive random memories (ReRAM) appear as promising candidates to overcome some of the above limitations. Briefly, ReRAM is a two terminal device where information is written by voltage pulses and stored in its resistance state. Binary ON/OFF states (high resistance (HRS) and low resistance (LRS)) can be stored. Information can be read by using the same two terminals and recording its resistance. ReRAM can be scaled down below 10 nm and offers a fast, non-volatile and low energy electric switching. Moreover, multiple resistance states of the device can be fixed depending on the device history (voltage-time excursions, $V(t)$), thus constituting a *memristo*r element.[3] There are a variety of resistive switching (RS) mechanisms leading to resistance change.[4] Among the most promising are those based on the so-called soft-dielectric breakdown mechanism (sDB), where a conducting path or filament (CF) is opened or closed (ON/OFF states) depending on the $V(t)$ signal applied to the device. The semiconductor industry has been using metal-oxide and metal-insulator-semiconductor junctions for CMOS technology since decades, and, maybe for this reason, resistive switching using oxides has been much explored. Alternative resistive switching approaches, such as phase change, appear to fall behind in the race due to the higher required power.[2]

$HfO_2$ is an excellent candidate to be integrated in ReRAM devices, showing, up to now, the best performance[5,6] and being easy to integrate in 3D crossbar arrays.[7] However, ReRAM devices suffer from an important variability due to the RS mechanism.[8] The switching mechanism in $HfO_2$ is believed to be related to a sDB and the formation of CFs. In $HfO_2$, CFs are understood as oxygen-deficient channels created by electric-field induced ionic migration enhanced by a Joule local heating.[9,10] Indeed, it is known that grain boundaries are preferential mobility paths for oxygen vacancies in $HfO_2$,[5,11] as well as in other functional oxides.[12–14]

The discovery of ferroelectricity (FE) in ultrathin films of Si and Al-doped $HfO_2$ in 2011[15,16] opened new avenues for this material.[17] Indeed, ferroelectric field transistors (Fe-FET) based on $HfO_2$ are now implemented in CMOS technologies,[18–20] opening the path to non-volatile Fe-FET memories based on FE $HfO_2$. Besides, ferroelectric materials offer other opportunities, such as tunnel barriers in metal-insulator-metal structures. The switching of the ferroelectric polarization direction changes the characteristics of the tunnel barrier (height and width) and, therefore, its conductance.[21] Data (polarization direction) can be subsequently read without changing the memory state and refreshing is no longer required. As the direction of polarization is non-volatile and can be written by a suitable voltage, FE tunnel barriers are promising materials for low-power ReRAM.[22] Consequently, ferroelectric $HfO_2$ films are being explored for ferroelectric tunnel devices (FTJ)[23] and negative capacitance gate insulators.[24]

From the material's point of view, a crucial aspect is that, whereas RS devices based on CFs can be operated with amorphous $HfO_x$, ferroelectric barriers require crystalline materials.

HfO$_2$ can crystalize in different polymorphs: monoclinic, tetragonal and cubic phases. By doping with atoms such as Si, Zr, Y, Gd, La, etc.,[17] HfO$_2$ can be stabilized in an orthorhombic phase, which is ferroelectric. Among the dopants, Zr allows the stabilization of the ferroelectric orthorhombic phase in a wider dopant concentration, being Hf$_{0.5}$Zr$_{0.5}$O$_2$ (HZO) the optimal composition.[25] Nanometric films of ferroelectric HZO have been grown by atomic layer deposition, sputtering and more recently, by pulsed-laser deposition. In general, the films display the coexistence of crystallites of o-HZO and monoclinic (m-HZO) and/or other phases. The growth of epitaxial HfO$_2$-based ferroelectrics on single crystalline substrates has been recently reported on YSZ,[26–30] on perovskites substrates[31–34] and on Si.[35–37] However, their resistive switching behavior has been scarcely studied. The electroresistance (*ER*), defined as the ratio between resistance in the ON/OFF states of a tunnel barrier in a metal-insulator-metal structure, is its primary figure of merit. Nevertheless, as early recognized by Max et al.,[38] *I-V* curves from a ferroelectric HfO$_2$-based tunnel barrier can include mixed up the RS response associated to the formation/destruction of CFs and to polarization reversal, challenging their understanding and control. In fact, *ER* in HZO tunnel barriers have been recently reported,[33,39–41] but the assignment of *ER* solely to the polarization reversal remains inconclusive. The difficulty originates from the fact that HZO films grown on single crystalline perovskite substrates, typically contain crystallites of the paraelectric m-HZO phase in addition to crystallites of the o-HZO phase. Estandía et al.[42] reported that HZO films grown on LaAlO$_3$ (LAO), SrTiO$_3$ (STO) and RScO$_3$ (R being a rare earth) single crystalline (001)-oriented substrates, buffered by a suitable oxidic metallic electrode (La$_{2/3}$Sr$_{1/3}$MnO$_3$, LSMO), display a systematic increase of the o-HZO/m-HZO ratio from LAO to STO and to RScO$_3$. Moreover, the ferroelectric o-HZO phase is found to be (111) oriented on the cubic (001) surface of the substrates and, therefore, o-HZO is intrinsically twinned.[42] It follows that grain boundaries between twinned o-HZO crystallites (denoted GB-I), as well as grain boundaries between o-HZO and m-HZO and m-HZO and m-HZO crystallites (denoted GB-II) should exist in the film, as sketched in Figure 1a. Notice that these GB can be coherent or not, depending on the nature of the crystallites involved and growth conditions.

The presence of these GBs may have a dramatic effect on the RS of the films because, as mentioned, GBs are highways for ionic motion. Indeed, recently M.C. Sulzbach et al.[43] identified coexistence of different RS mechanisms contributing to the *ER* of HZO-based ferroelectric tunnel junction. One contribution, occurring at low voltage which closely coincides with the coercive field of the ferroelectric ($V_C$), is associated to polarization reversal and, therefore, involves only the motion of polarization charges and displacive currents. A second contribution occurs at voltages above a threshold value ($V_{GB}$), which was assigned to field-induced ionic motion (Figure 1c). Interestingly, it was observed that the ionic contribution was prevalent in films containing mixed GB-I and GB-II grain boundaries, whereas the genuine polarization-related RS was prevalent in films containing only GB-I. The relative amount of each of the grain boundaries can be controlled with the substrate used in the heterostructure.[42]

A way to falsify the scenario described above is to block the charge leakage along GBs in HZO tunnel barriers. In this regard, the simplest approach is to *seal* the GBs with a suitable nanometric oxide layer, conformably capping the surface of the HZO films and the GBs embedded in them, and subsequently inspect their *ER*. It should be expected $V_{GB}$ to be shifted to higher voltages and eventually suppressed (Figure 1b,d), helping to enhance the robustness of the ferroelectric related *ER*.

Here, we first report the *ER* of thin HZO (4.6 nm) films grown on $SrTiO_3$ and $GdScO_3$ substrates, which allows controlling the relative abundance of GB-I and GB-II grain boundaries in the films.[42] Then, dielectric $AlO_x$ and $SrTiO_3$ layers of different thickness (1-2 nm range) were grown on HZO and the voltage-dependent *ER* of the junctions was measured. It will be shown that the dielectric layers indeed have a tremendous positive impact on device performance. They shift the onset of ionic-like RS, occurring at $V_{GB}$, towards higher voltages, thus stablishing a wider range of genuine tunnel ferroelectric-polarization controlled resistive switching behavior. In this way, a more robust operation than tunnel devices based on bare ferroelectric HZO barriers is obtained and, remarkably, the auxiliary dielectric capping layers allow to reduce by about a factor 7 the device-to-device standard deviation of *ER* values.

**Results and discussion**

Epitaxial HZO films of 4.6 nm nominal thickness were grown on SrTiO$_3$ (001) (STO) and GdScO$_3$ (001)-oriented (using pseudo cubic setting) (GSO) single crystalline substrates (5 x 5 mm$^2$) buffered with La$_{2/3}$Sr$_{1/3}$MnO$_3$ (22 nm thick) conducting electrodes by pulsed laser deposition (PLD), as described elsewhere.[42] HZO was subsequently capped with either STO or AlO$_x$ layers deposited by ablating SrTiO$_3$ and Al$_2$O$_3$ targets, at PO$_2$ = 0.02 mbar, T = 700ºC and cooled to room temperature under PO$_2$ = 0.2 mbar, and P(Ar) = 0.1 mbar at room temperature, respectively. The thickness of the STO and AlO$_x$ capping layers was chosen to be t = 0, 1 and 2 nm, controlled by the number of laser pulses on the basis of calibrated growth rates previously determined. Circular Pt top electrodes of 20 µm of diameter were grown through shadow masks by sputtering. The heterostructure is sketched in Figure 2a. The reference sample, i.e., in which no dielectric were deposited beween the the ferroelectric and platinum, is labelled as HZO. The samples with 1 or 2 nm STO capping are named: HZO/STO1 and HZO/STO2, respectively. In the case of 1 and 2 nm AlO$_x$ capping, samples are named: HZO/AlO1 and HZO/AlO2, respectively. The samples grown on GSO substrate are labelled with GSO// in front.

Figure 2b shows a X-ray diffraction 2θ-χ frame and the corresponding integrated θ-2θ scan of the HZO/STO1 heterostructure grown on STO. The intense (00l) reflections of substrate and LSMO can be observed, together with a strong (111) reflection of the ferroelectric o-HZO and a small and broad peak at m-HZO(002) position, indicating that o-HZO is prevalent in the HZO film. In the right pannel, a zoom in the range 25-40° provides a better view of the o-HZO(111) and m-HZO(002) peaks. Pole figures indicate the existence of three different variants of the [111] textured HZO crystallites.[42] 2θ-χ frames collected on the bare HZO structure and HZO/STO2 films are also virtually identical (Figure S1). Similarly, the X-ray diffraction patterns of HZO/AlO1 and HZO/AlO2 do not reveal any difference with the bare HZO heterostructure (Figue S2).

Illustrative *I-V* curves recorded at 5kHz using PUND technique on HZO, HZO/STO1 and HZO/AlO1 devices are shown in Figure 2c. The corresponding *P(V)* loop shown in Figure 2d assesses the ferroelectric nature of the film. The coercive voltage extracted from the position of the switching peaks of the HZO sample is $V_C^+ \approx V_C^- \approx 3$ V. Data indicates a remnant polarization of $P_R \approx 14$ μC·cm$^{-2}$. These ferroelectric characteristics are consistent with previous reports.[34,43,44] A minor modification of the shape of the I-V curves is observed in HZO/STO1 compared with the uncapped sample. In the case of HZO/STO2, the coercive field is slightly smaller and the polarization is reduced (Figure S3). One might ascribe the reduction of the coercive field to the STO large polarizability contributing at lower fields. Furthermore, in HZO/AlO1 and HZO/AlO2 (Figure 1c and Figure S3) the switching peak is reduced and shifted to a slightly larger voltage without important variations between both AlO thicknesses. These results are expected[45] due to the better insulating properties and the lower permittivity of AlO$_x$ compared with STO.

At this point, having established the ferroelectric nature of all HZO films in HZO/capping structures, one can proceed to determine their *ER*. To measure *ER*, first a writing trapezoid pulse with amplitude $V_W$ and duration $\tau_w = 300$ μs is applied using the electrical contact configuration shown in Figure 2a. In order to avoid ferroelectric switching during the reading, the maximum reading voltage must be smaller than $V_C$ (≤ 2,5 V). I-V curves are collected, after a delay time $\tau_D = 0.5$ s, by applying a linear $V_R(t)$ pulse in a small voltage range (from -1 V to +1 V). The resistance is determined at $V_R = 0.9$ V.

Figure 3a shows the electric resistance ($R$) of a HZO junction measured using the writing voltage amplitudes ($V_W$) indicated. It is clear that $R(V_W)$ displays two well defined regions, labelled I and II. As discussed in M.C. Sulzbach et al,[43] the electroresistance in these regions has different origins. In region I, $ER$ is mainly dictated by the polarization reversal which changes the characteristics of the HZO tunnel barrier and thus the carrier transport across it. Indeed, above a certain voltage threshold (around 2 V), $V_W > 0$ and $V_W < 0$ dictate an increase or reduction of the junction resistance, as typically observed in ferroelectric tunnel junctions.[46,47] However, when entering region II, at $V_{GB} \approx 5$ V in Figure 3a, a new conducting channel opens and the junction resistance decays for both polarities. M. C. Sulzbach et al.[43] concluded that $V_{GB}$ is a fingerprint of the onset of non-polarization related conducting paths along grain boundaries.[8] Indeed, grain boundaries between the majority ferroelectric orthorhombic o-HZO phase and the non-ferroelectric monoclinic phase (m-HZO) were observed by means of STEM.[42] It was found that $V_{GB}$ shifts to higher voltages and eventually disappears when suppressing the occurrence of m-HZO grains by appropriate substrate selection for the HZO growth.

A major finding is presented in Figure 3b and 3c, where the $R(V_W)$ data collected in junctions containing the STO capping layer are depicted. It is obvious that the capping STO layer delays the entrance to region II, i.e $V_{GB}$ increases. Indeed, $V_{GB}$ (HZO/STO1 and HZO/STO2) is about 6 V, which is 20% larger than $V_{GB}$ (HZO) $\approx 5$ V. The observation of a wider region I in STO-capped HZO layers implies that the resistance contrast ($\Delta R = R(V_W^+)-R(V_W^-)$) in region I can be increased by increasing $V_W$ until full polarization saturation of the ferroelectric layer (dashed line), without ionic-like conducting channels. As shown in Figure 3a-c, $ER$ at the maximum $V_W$ ($< V_{GB}$) for HZO/STO1 is not significantly larger (390%) from that observed in HZO samples (340 %). On the other hand, it is observed that, irrespectively of the polarity of $V_W$, the resistance of the junctions increases with the capping layer (series resistance), particularly noticeable at HZO/STO2 (Figure 3c), overall limiting the observed

ER. Similar increase of $ER$ with $V_W$ is obtained in most of the HZO/STO1 and HZO/STO2 measured junctions (Figure S4).

An identical set of measurements has been performed on samples capped with $AlO_x$. Data for some illustrative junctions are shown in Figure 3(e,f). An increase of $V_{GB}$ and the subsequent expansion of region I can be observed in $AlO_x$ capped samples. The best improvement is observed for HZO/AlO1, in which the electroresistance is $ER$ ($\approx$ 700 %). The results reported in Figure 3(a,f) strongly suggest that, upon capping the HZO film with STO or $AlO_x$, grain boundaries within the HZO film have become inactive to open CFs. GBs in HZO films can be coherent or incoherent, depending on the crystallites involved. The ferroelectric o-HZO phase, which is (111) textured on cubic (001) substrates, must display threefold twin planes, thus, giving rise to a network of coherent GBs among them (GB-I). Nevertheless, the residual presence of the m-HZO phase in films grown on $SrTiO_3$ (001) implies that incoherent GBs between the m-HZO and o-HZO phase should also exist (GB-II).

The results shown in Figure 3 indicate that the capping layers have a major effect on the GB-II by cancelling or retarding the formation of conducting channels across the film. The reason why the conducting channels are not completely cancelled, particularly in films grown on STO, might result from the granular character of the STO and AlO capping layers (Figure S5), precluding full efficiency on blocking all ionic chanels along GB. Therefore, it should be expected the capping layers to have a smaller effect in HZO films where only the o-HZO phase exists and thus GB-II are absent. As mentioned, it has been demonstrated[42] that epitaxial HZO films on scandate substrates ($GdScO_3$ or $TbScO_3$, for example) yields a majority of o-HZO ferroelectric grains and their associated GBs are GB-I, pushing towards larger $V_W$ voltages the entrance into region II.

Therefore, to validate the hypothesis that the capping layer blocks mainly charge transport along incoherent grain boundaries, one should compare the electroresistance of GSO//HZO and GSO//HZO/STO2 heterostructures.

Accordingly, a 2 nm thick STO capping layer was grown to form GSO//HZO/STO2 junctions and its electroresistance was compared to uncapped GSO//HZO devices. Figure 4a, shows the *I-V* curves measured at 5kHz using PUND on GSO//HZO and GSO//HZO/STO2 devices. The *I-V* curve of GSO//HZO structures displays regular, abrupt and symmetric ferroelectric switching current peaks. As already observed in Figure 2, STO capping reduces the switchable polarization and only marginally deforms the current switching peak.

Figure 4b shows the $R(V_W)$ of a representative junction on GSO//HZO measured using the same protocol as in Figure 3. It is observed that $V_{GB}$ occurs at large voltage ($V_{GB} \approx$ 12-13 V), sharply contrasting with $V_{GB} \approx$ 5 V observed in HZO grown in STO substrates (see Figure 3c). This tremendous voltage shift is a fingerprint of the absence of voltage-induced conducting channels across GB-II grain boundaries in GSO//HZO samples, as already reported.[43] The relevance here is that the GSO//HZO/STO2 shows stable ER in an expanded voltage range, up to the largest explored $V_W$ (15 V). This indicates that these GSO//HZO/STO2 junctions can be operated up to 15 V (Figure 4b) without any significant contribution of GBs to the ER, and that the STO dielectric layer cancels any residual contribution of ionic conduction that may exist accross GB-I grain boundaries. This observation provides a hint that ionic conductance accros GB-I, which exists in the film due to its twinnig origin, is more limited than accross GB-II. Notice that the forming step, which opens conductive filaments in conventional ferroelectric $HfO_2$-based devices, is in the few volts range.[38] It follows that the electroresistance of FTJs on HZO should be more stable and robust if a capping layer is used. This claim is illustrated by showing in Figure 5(a-c) the spread of ON/OFF resistance values collected from HZO junctions with mean value of $3.4 \times 10^9$ Ω and $3.3 \times 10^8$ Ω with standard deviation of $2.6 \times 10^9$ Ω and $3.8 \times 10^8$ Ω for OFF and ON, respectively. Data in Figures 5(b,c), corresponding to HZO/STO1 displays an obvious improvement when compared to the bare HZO junctions. For HZO/STO1 the mean value is $1.5 \times 10^{10}$ Ω and $5.2 \times 10^9$ Ω with standard deviation of $1.4 \times 10^{10}$ Ω and $1.5 \times 10^9$ Ω for OFF and ON, respectively. In the case of HZO/AlO1, the mean value is $8 \times 10^9$ Ω and $1.3 \times 10^9$ Ω with standard deviation of $3.3 \times 10^9$ Ω and $8.3 \times 10^8$ Ω for OFF and ON, respectively. The spread of the ER values evaluated from its standard deviation [σ (ER)] reduces by about a factor 7 when comparing data for bare HZO (σ (ER) ≈ 1400%) and HZO/STO1 junctions ((σ (ER) ≈ 113%). The relative change of σ(ER) (%) (= σ(ER)/mean(ER)) is also improved, although to minor extent (roughly a factor 3) (for instance: 56% STO capped, 114% bare HZO) due to the fact that overall resistance of the junctions increases with capping thus reducing to ER. The capping layers can also promote a higher endurance of the *ER* response. To address this issue, repeated $ER(V_W)$ loops have been performed on a given junction, either on bare or on capped HZO films. Some illustrative data are displayed in Figure 5(d-f). Data in Figure 5(g-i) provide an even stronger evidence of this assessment. Indeed, it can be appreciated in Figure 5h that the *ER* of the HZO/STO1 junctions have a robust *ER* that can be cycled at least up 300 times., largely contrasting with the poorer performance of the bare HZO junctions (Figure 5g).

In order to get further information about the role of the capping layer on the performance of the HZO/(STO, AlO) heterostructures, their miscrostructure were studied by means of STEM in combination with electron energy loss spectroscopy (EELS). In **Figure 6**a (main panel) a High Angle Annular Dark Field (HAADF) image of the HZO/STO2 heterostructure is shown. The HAADF imaging mode produces Z-contrast images which allows for recognizing the different layers of the heterostructure according to the atomic number Z of their constituent elements. The excellent quality of the bottom electrode, the crystalline nature of the HZO layer, conformally covered by a polycrystalline STO capping layer and the Pt electrode can be clearly recognized. In the selected field-of-view of Figure 6a, crystallites of the (111) textured o-HZO phase can be identified. As mentioned, the growth of (111)-HZO textured on (001) cubic substrates implies the existence of twined HZO crystals,[42] which can be observed in the enlarged view in Figure 6b. The different orientation of the HZO crystallites, indicated by the corresponding zone axis, yields grain boundaries of GB-I type. As observed in the image, the STO capping layer has grown well crystallized on the HZO layer and continuously covers the GB-I between o-HZO crystallites. Naturally, in other regions of the sample, type GB-II are found and similarly, the STO layer covers the grain boundary region. In Figure S6, illustrative examples of GB-I and GB-II grain boundaries observed in HZO/STO1 are shown. Further proof of the sealing of the HZO's grain boundaries comes from EEL spectrum imaging. Figure 6c (top panel) shows the HAADF image acquired simultaneously to EELS data. Figure 6c (bottom panel) displays the corresponding EELS elemental maps obtained from Pt-M (in red), Ti-L (in blue) and Zr-L (in yellow) edges. The elemental maps show that the different layers are continuous (see Figure S7 for a larger field of view all the elemental maps), and therefore, that the STO capping layer constitutes an effective barrier between Pt and HZO. In conclusion, GB-I and GB-II are covered by the 2 nm STO layer.

To get an insight on a possible chemical role of the capping barrier on the HZO on the distinctive *ER* of capped and uncapped HZO devices, the fine structure of both the O-K and the Ti-L edges were analyzed (see Figure S7) and their profiles in HZO (capped and uncapped) and STO (capping layer and substrate) compared, respectively. It has been found that the O K scans do not allow to discern any change in uncapped and capped HZO. In contrast, in the STO capping layer, a remarkable shift of the Ti-L edge has been observed indicating a reduction of the valence state of $Ti^{m+}$. More precisely, the observed energy shift would correspond to a nominal $Ti^{3+}$ state, sharply contrasting with the expected $Ti^{4+}$, as observed in the substrate. Although dedicated studies are required to fully address this observation, it suggests that the capping layer may act as an oxygen reservoir to the HZO getter, perhaps contributing to the sealing of grain boundaries in HZO by oxygen-vacancy filling.

Before closing, it should be mentioned that recently Max et al.[23,38] reported an earlier attempt to introduce an AlO$_x$ layer in ferroelectric HZO devices. A nanometric AlO$_x$ layer was grown on symmetric metal-HZO-metal polycrystalline structures, in order to break the symmetry of the junction and to exploit the tunnel transport through the AlO$_x$ barrier. The approach here is radically different: advantage is taken from the fact that the polarization reversal of the ferroelectric barrier itself does change the charge transport. Although it cannot be excluded that in here, the dielectric layer STO or AlO$_x$ may have an additional effect related to changing of the energy profile of the barrier. Data show that the most prominent effect is blocking of the ionic channel.

**Summary and conclusion**

In HfO$_2$-based nanometric ferroelectric layers, resistive switching between high resistance (HRS) and low resistance (LRS) states determines the binary OFF/ON states. Resistive switching may not only be achieved by polarization reversal of the ferroelectric domains but also be contributed by the formation of conducting filaments across the HZO layer, the latter being mostly controlled by the presence of grain boundaries. In epitaxial films of HZO, grain boundaries between twinned o-HZO ferroelectric crystallites and grain boundaries between o-HZO and a residual paraelectric m-HZO phase may create a grain boundary network. The presence of these grain boundaries sets an upper voltage limit for device operation in the reversible and genuine ferroelectric switching regime, in which displacive currents rather than transport charges flow occurs with the concomitant benefit of reduced Joule losses and endurance. Here, it has been shown that capping the functional HZO layer by ultrathin crystalline SrTiO$_3$ or amorphous AlO$_x$ layers allows for a radical improvement of the operation window of the tunnel device in the ferroelectric region. It has been observed that these capping layers allow to increase the yield and homogeneity of functioning ferroelectric junctions and to strengthen their endurance. STEM images have shown that a polycrystalline

STO coating is indeed conformal. However, crystallinity is not seen to be essential in the observed upgrade of the electroresistive response given that a similar effect is observed in amorphous $AlO_x$ capping. Therefore, the ultimate reasons for the observed beneficial effects of the capping remain to be disclosed. The observation that Ti reduction occurs in the STO capping layer may suggest that this layer acts as an oxygen source for the grain boundaries of HZO. In this scenario, it could be envisaged to explore and test other oxides for fully compatible CMOS technologies.

**Experimental**

*Sample Growth:* Epitaxial HZO films with nominal thicknesses 4.6 nm were grown on LSMO (22 nm thick) buffered (001) STO and (001)-oriented (pseudocubic indexation) GSO as described elsewhere.[42] The dielectrics STO and $AlO_x$ were grown in situ at 700°C and room temperature, respectively. Pt electrodes (20 nm thick) were grown ex-situ, at room temperature, by DC-sputtering through shadow mask which allows the deposition of circular top contacts (diameter ≈ 20 μm) in all samples. Single crystalline STO and GSO used as substrates have bulk cell parameters (pseudo-cubic) of 3.905 Å and 3.97 Å, respectively. Full structural characterization of HZO films on those substrates has been reported elsewhere.[42]

*Structural Characterization:* X-ray diffraction 2θ-χ recorded using Bruker Bruker-AXS D8 Advance equipped with a Vantec 500 detector (Cu-K$_α$ radiation). The corresponding θ-2θ scan was obtained by integration within the ± 10º range angular.

*Electrical Characterization*: The electrical contact configuration is sketched in Figure 2a. The bottom LSMO layer acts as electrical ground and it was contacted through silver paste contact at the edge of the sample. Top Pt electrodes were biased. The *I-V* curves were measured using the PUND technique.[48,49] Polarization measurements were done by integrating the current though time of collected *I-V* curves at 5 kHz. Electroresistance measurements are done by using a trapezoid voltage pulse of variable amplitude $V_W(t)$ and duration $τ_W$ (300 μs) to set the device initial state (writing step). A linear $V_R(t)$ pulse of maximum amplitude 1 V is subsequently used to read the resistance of the junction. Quoted values of junction resistance correspond to $V_R$ = 0.9 V. The delay time $\tau_D$ is fixed at 0.5 s. All electrical characterizations were performed with an AixACCT TFAnalyser2000 platform.

*Scanning Transmission Electron Microscopy Characterization.* Aberration-corrected scanning transmission electron microscopy (STEM) was used for microstructural analysis with atomic resolution. Samples were characterized using a JEOL JEM ARM200cF operated at 200 kV, equipped with a CEOS aberration corrector and GIF Quantum ER spectrometer, at the Universidad Complutense de Madrid, Spain. The STEM images were acquired in high angle annular dark field imaging mode, also referred to as Z-contrast because the brightness associated to each atomic column roughly scales with the square of the atomic number Z.[50] The STEM specimens were prepared using a FEI Helios nanolab 650.

**Acknowledgements**

Financial support from the Spanish Ministry of Economy, Competitiveness and Universities, through the "Severo Ochoa" Programme for Centres of Excellence in R&D (SEV-2015-0496) and the MAT2017-85232-R (AEI/FEDER, EU), and MAT2015-73839-JIN projects, and from Generalitat de Catalunya (2017 SGR 1377) is acknowledged. MC acknowledges fellowship from "la Caixa Foundation" (ID 100010434; LCF/BQ/IN17/11620051). IF and JG acknowledge Ramón y Cajal contracts RYC-2017-22531 and RYC-2012-11709, respectively. SE acknowledges the Spanish Ministry of Economy, Competitiveness and Universities for his PhD contract (SEV-2015-0496-16-3). MC work has been done as a part of her Ph.D. program in Physics at Universitat Autònoma de Barcelona. SE work has been done as a part of his Ph.D. program in Materials Science at Universitat Autònoma de Barcelona. Authors acknowledge the SEM-FIB microscopy service of the Universidad de Málaga and the ICTS-CNME for offering access to their instruments and expertise. Huan Tan is aknowledged for performing the AFM characterization.

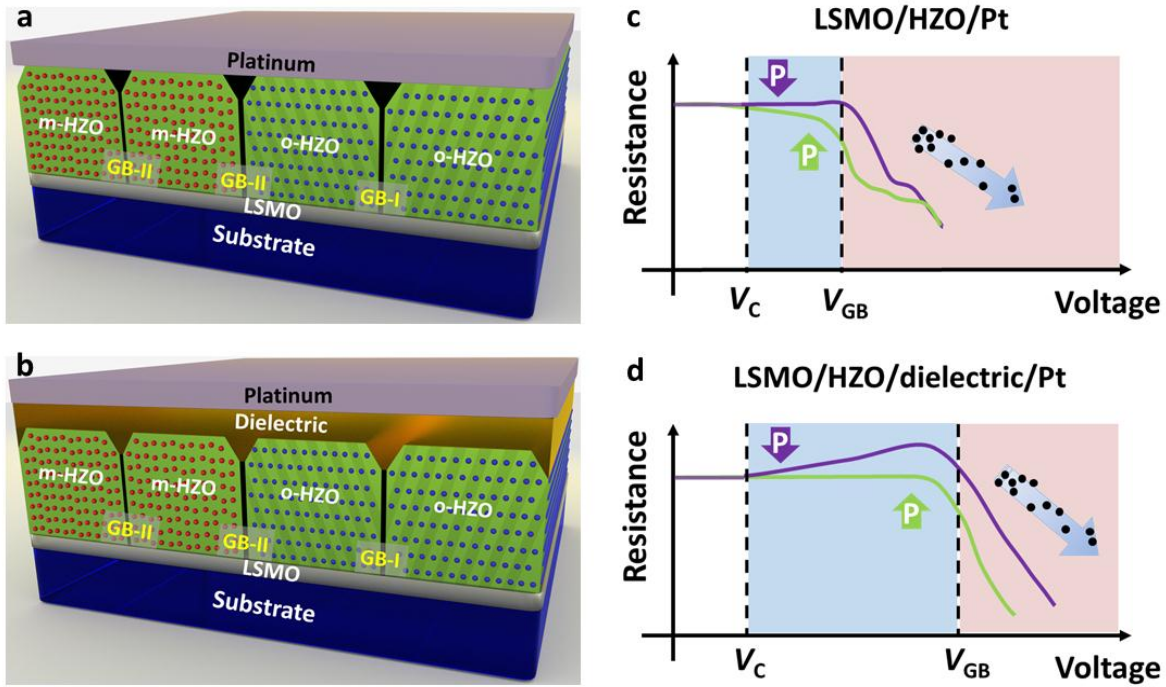

**Figure 1**. Sketch of epitaxial ferroelectric HZO films grown on $SrTiO_3$//LSMO containing o- and m-HZO grains, few nanometers in size, and grain boundaries (GB-I and GB-II) among them. (a) HZO film is covered by a metallic (Pt) electrode. (b) A dielectric layer is inserted between the HZO film and the metallic Pt electrode.(c) Sketch of the dependence of the resistance of ferroelectric HZO barrier on the writing voltage using the Pt electrode on the bare HZO surface as in (a). The purple curve represents values of R when positive bias is applied at the Pt electrode (P pointing to LSMO) and the green curve represents values of R when negative bias is applied (P pointing to Pt). (d) Sketch of the dependence of the resistance of ferroelectric HZO barrier on the writing voltage using a dielectric layer inserted between the electrode and the HZO layer as in (b). The voltages $V_C$ and $V_{GB}$ correpond to the coercive field of the ferroelectric and the opening of ionic-like conducting channels across the barrier, respectively. Up-down arrows indicate the polarization direction and dots-tilted arrow suggest ionic motion.

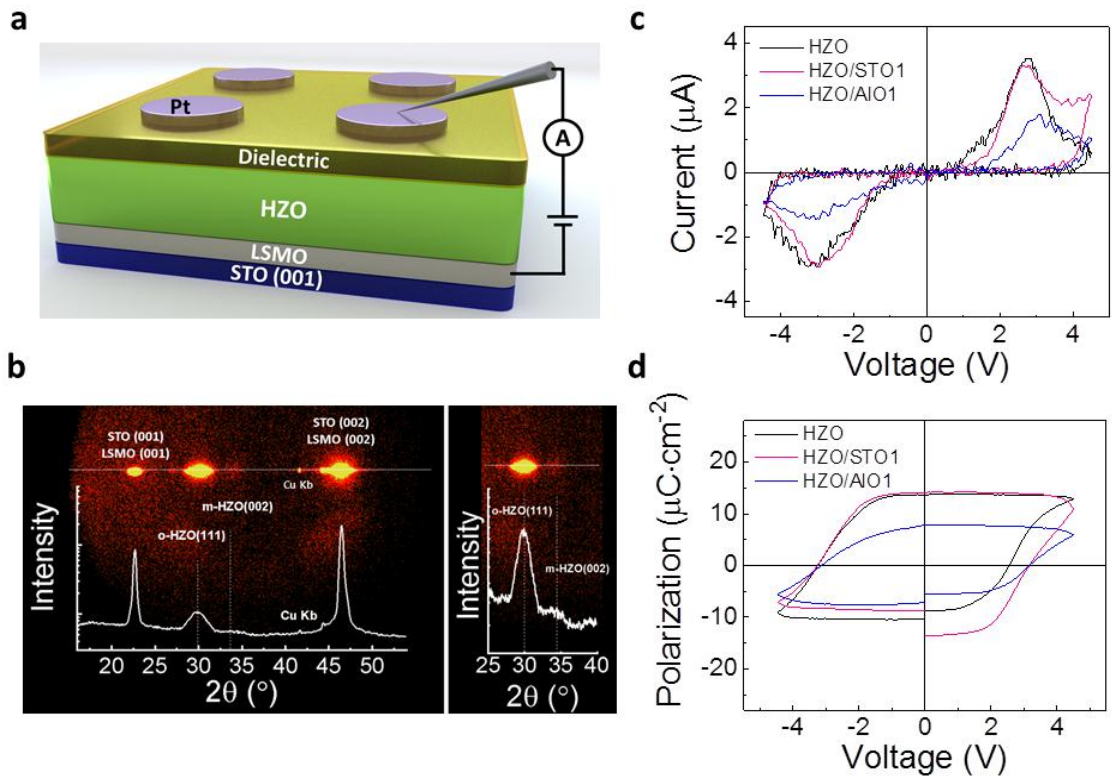

**Figure 2.** (a) Sketch of the sample heterostructure and electric contact arrangement. (b) Left: X-ray diffraction 2θ-χ frame of HZO/STO1 grown on STO substrate and the corresponding θ-2θ pattern. The (001)STO, (00l)LSMO, (111)o-HZO and (002)m-HZO reflections are indicated. Right panel: zoom of 25-40° range in which m-HZO(002) peak is visible. (c) Current-voltage loops ($V_{max}$ = 4.5 V) of junctions with platinum, STO (1nm) and AlO$_x$ (1nm) capping layer recorded at 5kHz. (d) The corresponding polarization $P(V)$ loop obtained integrating the I-V loops.

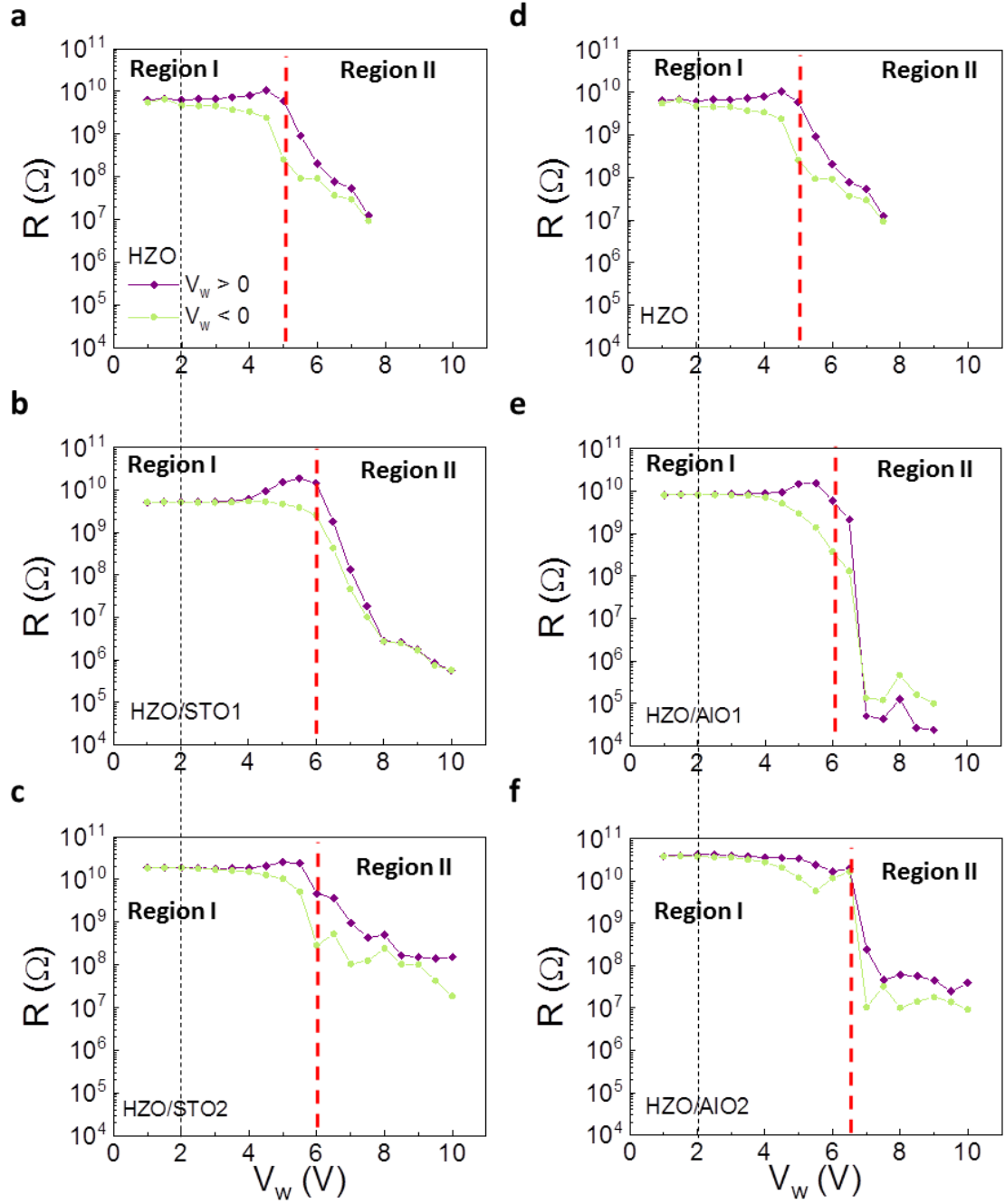

**Figure 3**. Resistance (at 0.9 V) of junctions: (a,d) HZO, (b) HZO/STO1 and (c) HZO/STO2 samples after positive ($V_W > 0$, purple) and negative ($V_W < 0$, green) writing pulses. (e,f) Equivalent data for (e) HZO/AlO1 and (f) HZO/AlO2 devices.

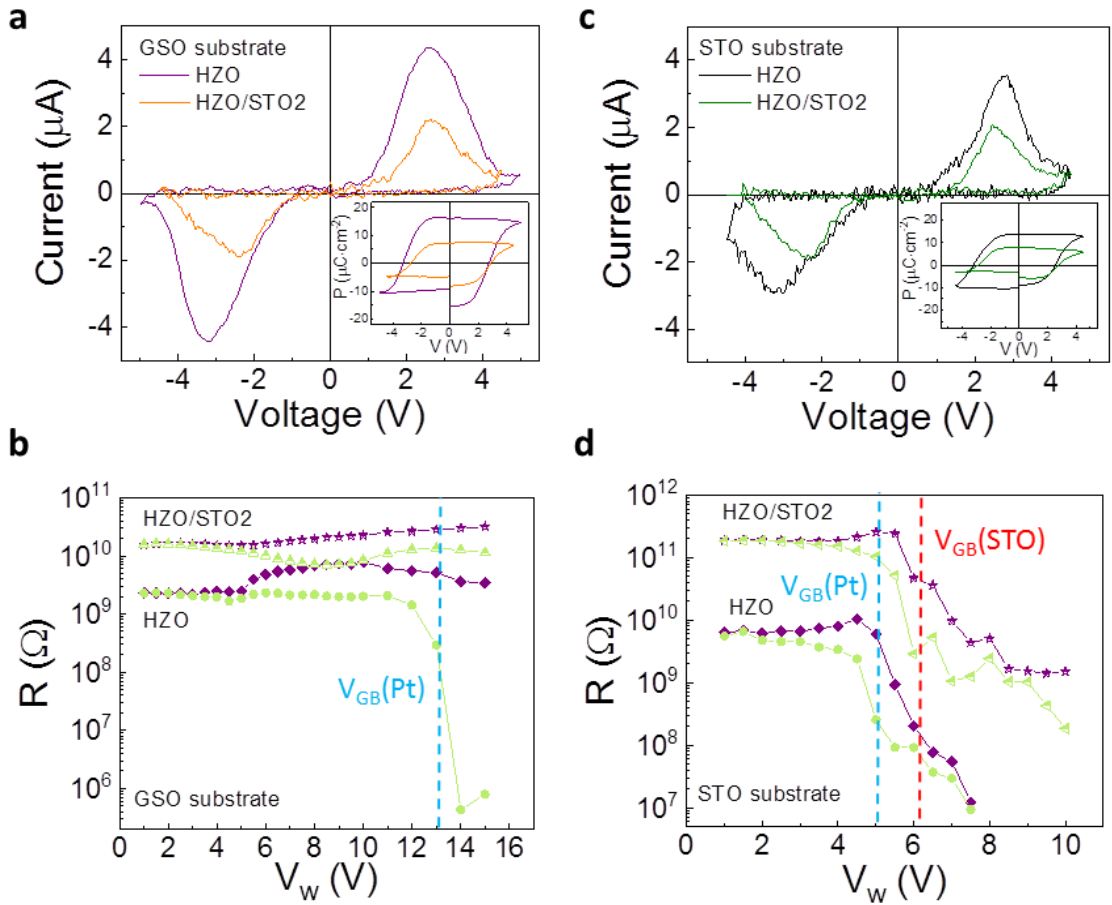

**Figure 4.** (a) I-V loop recorded at 5 kHz and (b) resistance $R(V_W)$ (after $V_W = \pm 5$ V) recorded on samples GSO//HZO and GSO//HZO/STO2. (c,d) The corresponding data for HZO and HZO/STO2 junctions (adapted from Figure 1 and 2) grown on STO substrate. In (d), for clarity, data for STO capping have been multiplied by 10. Vertical dashed lines in (b,d) indicate $V_{GB}$. Insets in (a) and (c) are the corresponding polarization loops.

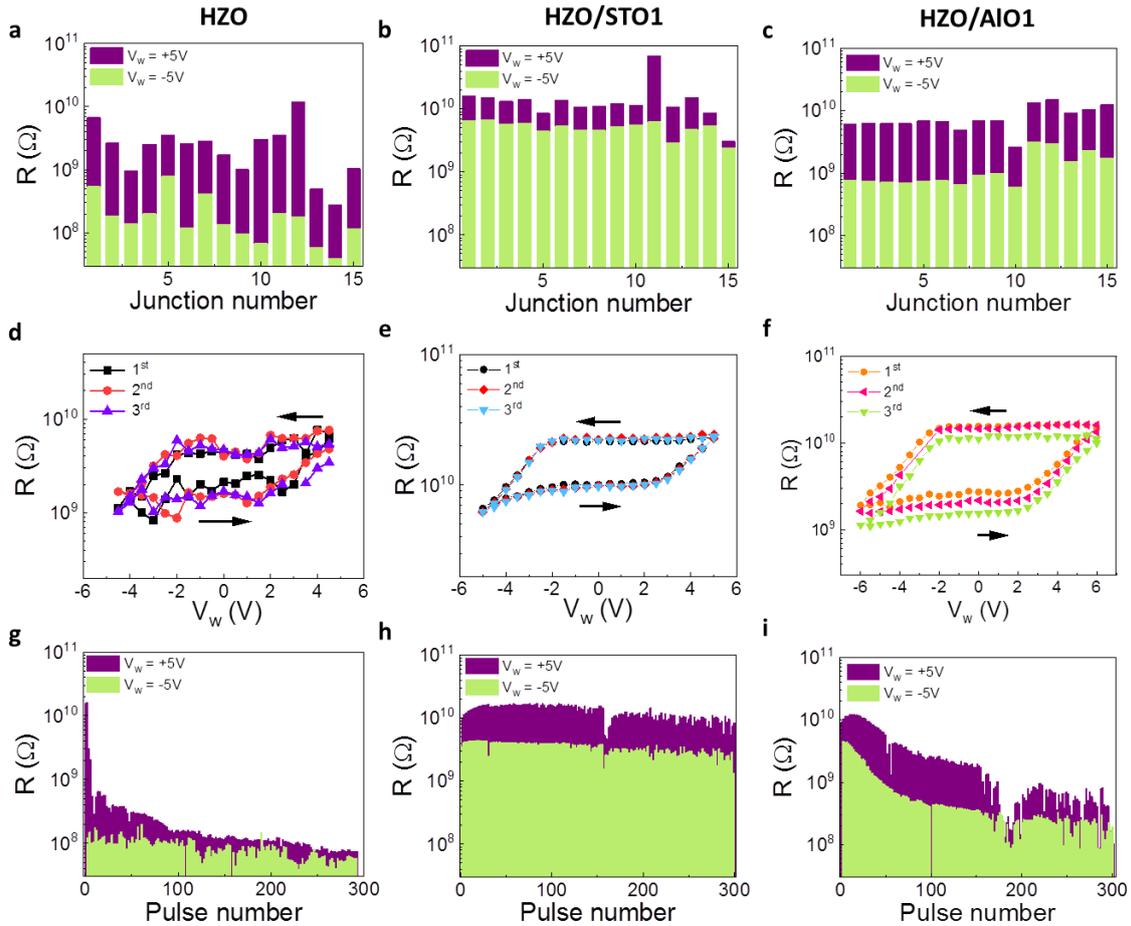

**Figure 5**. Electroresistance of HZO, HZO/STO1 and HZO/AlO1. (a-c) Resistance values recorded after writing with $V_W = \pm 5$ V on a set of 15 junctions on each sample. (d-e) Electroresistance $ER(V_W)$ loops recorded consecutively on uncapped and capped samples as indicated. (g-i) Endurance of the junctions for 300 pulses of $V_W = \pm 5$ V and $\tau_D = 300$ μs. After each writing pulse the resistance was measured at V = 0.9 V.

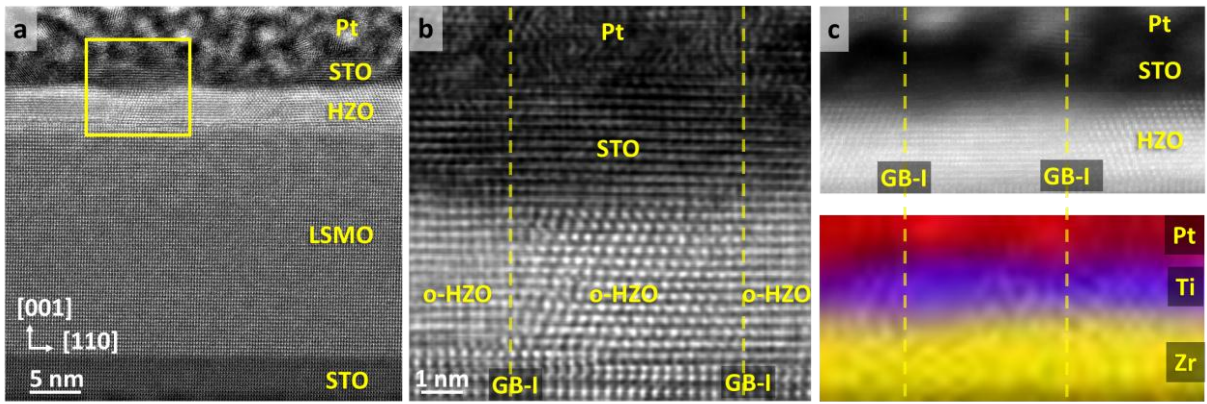

**Figure 6**. (a) Z-contrast image of a HZO/STO2 heterostructure viewed along the [110] STO substrate zone axis. STO substrate, LSMO and HZO layers, STO capping and the Pt layer can be distinguished. (b) Zoom of region marked in yellow in (a) where two GB-I and the STO capping layer fully covering the HZO film are visible. (c) The top panel shows the ADF image acquired simultaneously with the electron EEL spectrum image. The bottom panel shows an elemental map obtained from Pt-M (in red), Ti-L (in blue) and Zr-L3 (in yellow) edges. The EELS data show a conformal coating of Ti (and thus, STO) on HZO. Vertical dashed lines indicate the position of GB-I grain boundaries.

# Supporting Information

# Blocking of conducting channels widens window for ferroelectric resistive switching in interface-engineered Hf$_{0.5}$Zr$_{0.5}$O$_2$ tunnel devices

Milena Cervo Sulzbach, Saúl Estandía, Jaume Gàzquez, Florencio Sánchez, Ignasi Fina*, Josep Fontcuberta*

Institut de Ciència de Materials de Barcelona (ICMAB-CSIC), Campus UAB, Bellaterra, Catalonia 08193, Spain

**Supporting Information S1.**

Figure S1 shows X-ray diffraction 2θ-χ scans of HZO samples grown on STO substrates and capped with Pt, and STO (1 nm and 2 nm), and the corresponding θ-2θ scans. The intense (00l) reflections of substrate and LSMO can be observed as well as the (111) reflections from o-HZO and (002) reflections from m-HZO in all cases. It can be seen that the deposition of the top layers, either platinum or STO, does not affect the crystallinity of the HZO film.

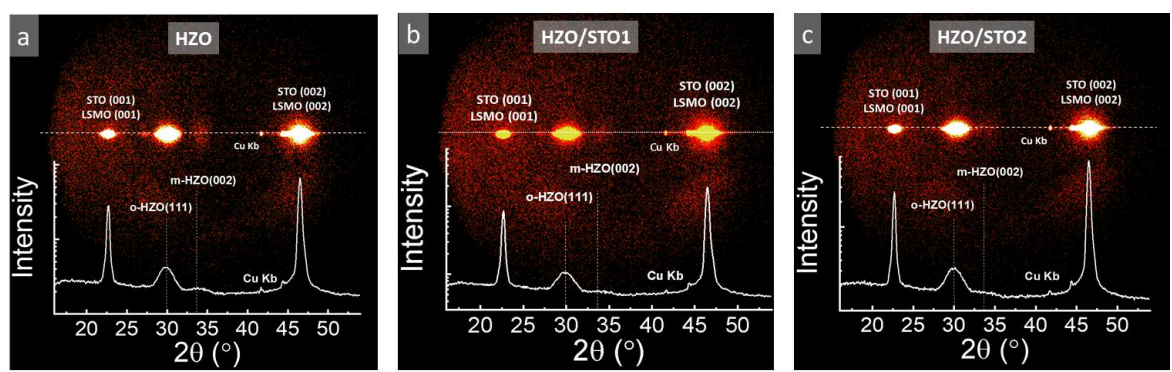

**Figure S1.** X-ray diffraction 2θ-χ scans (a) HZO, (b) HZO/STO1, and (c) HZO/STO2 films. The corresponding θ-2θ scans are also included.

**Supporting Information S2.**

Figure S2 shows X-ray diffraction 2θ-χ scans of HZO samples grown on STO substrates and capped with Pt, and AlO$_x$ (1 nm and 2 nm), and the corresponding θ-2θ scans. The intense (00l) reflections of substrate and LSMO can be observed as well as the (111) reflections from o-HZO and (002) reflections from m-HZO in all cases. It can be seen that the deposition of the top layers, either platinum or AlO$_x$, does not affect the crystallinity of the HZO film.

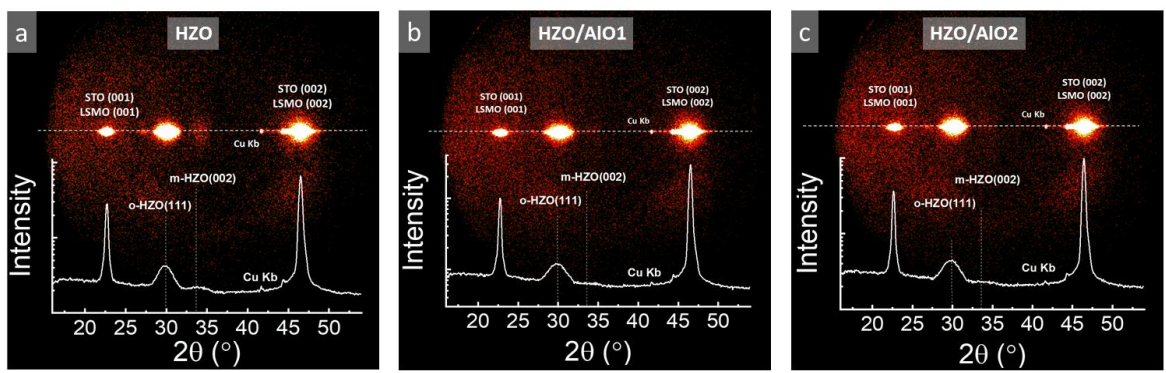

**Figure S2.** X-ray diffraction 2θ-χ frame of (a) HZO, (b) HZO/AlO1, and (c) HZO/AlO2 films. The corresponding θ-2θ scans are also included.

**Supporting Information S3.**

Figure S3 shows the current-voltage (I-V) loops recorded in HZO, HZO/STO2 and HZO/AlO2 junctions and their associated polarization curves. Reduction of switching peak amplitude for $t_{STO, AlOx}$ = 2nm is in agreement with reduction observed for 1nm, as it can be appreciated in Figure 2c in the main text.

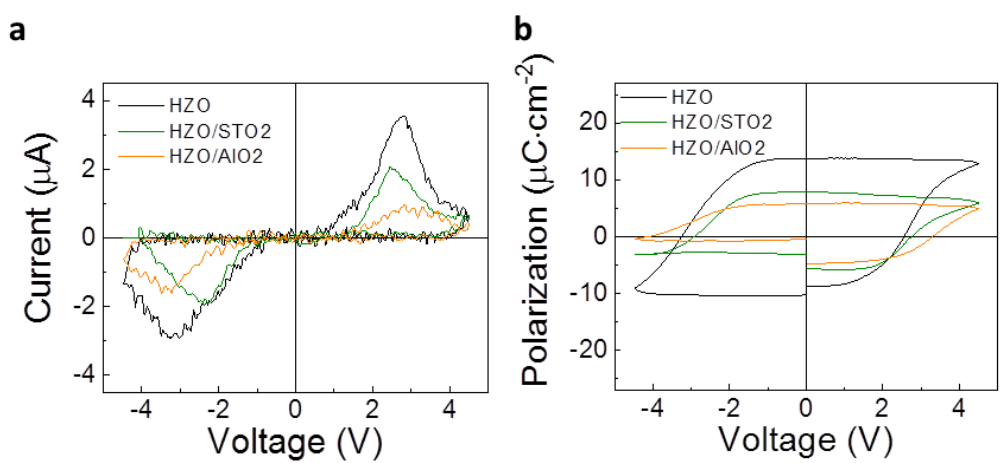

**Figure S3.** (a) Current-voltage loops ($V_{max}$ = 4.5 V) of HZO, HZO/STO2 and HZO/AlO2 junctions recorded at 5 kHz. (b) Polarization curves obtained by integration of I-V in (a).

**Supporting Information S4.**

The dependence of the electroresistance (*ER*) on the writing voltage ($V_w$) for a set of junctions is shown in Figure S4. As the writing voltage amplitude increases, the dispersion of results shrinks. Close to $V_{GB}$, the absolute *ER* value increases. For higher voltages, the resistance of the system drops and the junctions enter into Region II, associated to ionic-motion (see main text for details).

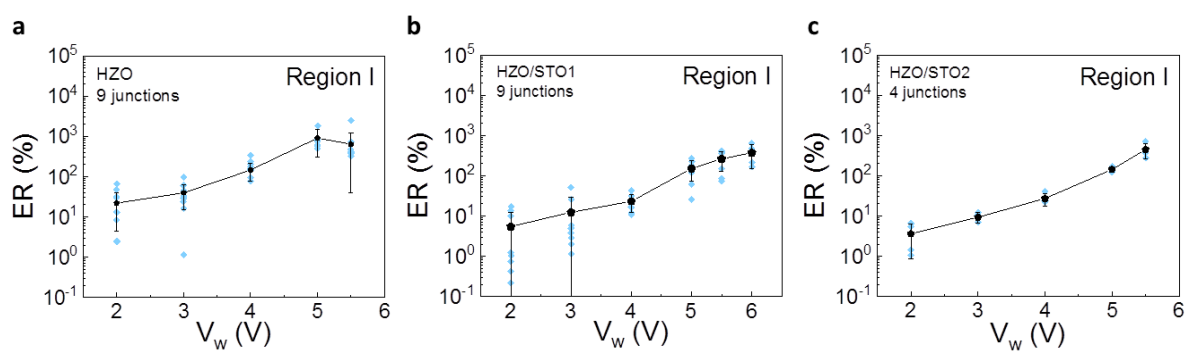

**Figure S4.** ER dependence on the writing amplitude ($V_w$) for (a) HZO, (b) HZO/STO1 and (c) HZO/STO2. The number of junctions measured in each sample are indicated in the graph.

**Supporting Information S5.**

Figure S5 shows topography images from atomic force microscopy (AFM) obtained at the surface of samples grown on STO substrates. Images from HZO/STO2 and HZO/AlO2 samples reflect the granular character of the capping layers. Roughness mean square values are 0.3, 0.3 and 0.2 nm for HZO, HZO/STO2, and HZO/AlO2 samples, respectively.

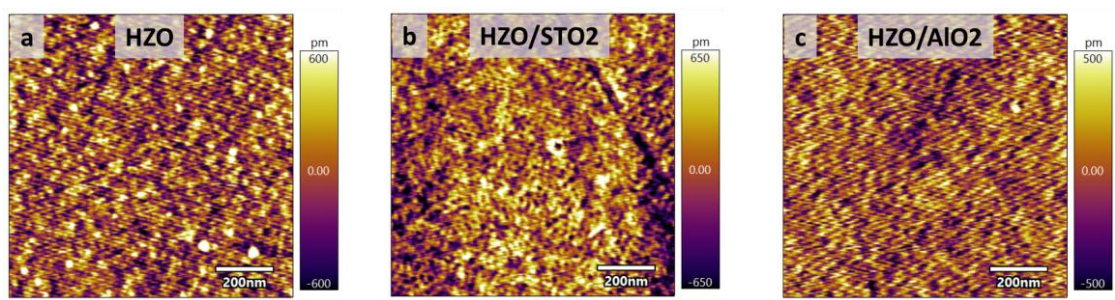

**Figure S5.** AFM topographic images of the surface of (a) HZO, (b) HZO/STO2 and (c) HZO/AlO2 samples. RSM values, excluding contamination, are 0.3, 0.3 and 0.2 nm, respectively.

**Supporting Information S6.**

Figure S6 illustrates the existence of GB-I and GB-II in the HZO film, well visible in the observed region in HZO/STO1.

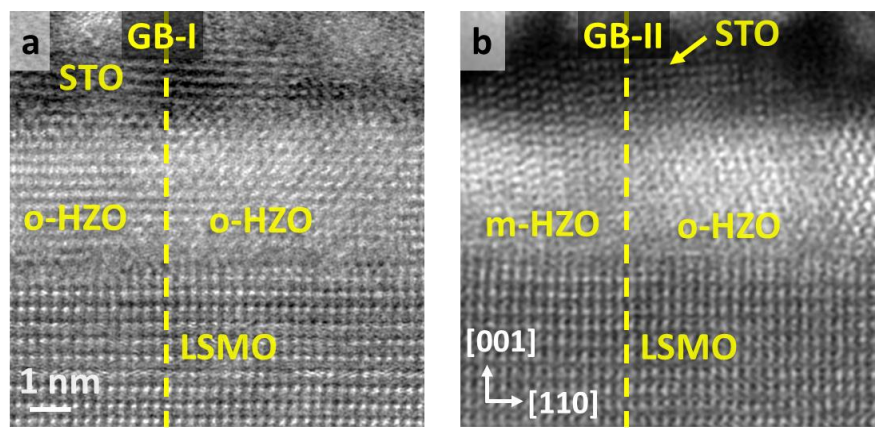

**Figure S5** (a) Z-contrast image of a HZO/STO1 heterostructure viewed along the [110] STO substrate zone axis. STO substrate, LSMO and HZO layers, STO capping and the Pt layer can be distinguished. Grain boundaries GB-I and GB-II can be well identified in (a) and (b) respectively.

**Supporting Information S7.**

Figure S7(a,b) shows an extended view of cross-section image of the HZO/STO2 sample, of Figures 6 (main text). The right panels of (b) shows the O-K edge (top) EELS spectra collected from two regions of the HZO layer in a HZO/STO capped asample as indicated; in the bottom panel we show Ti-L edge scans in the capping STO layer and STO substrate as indicated.

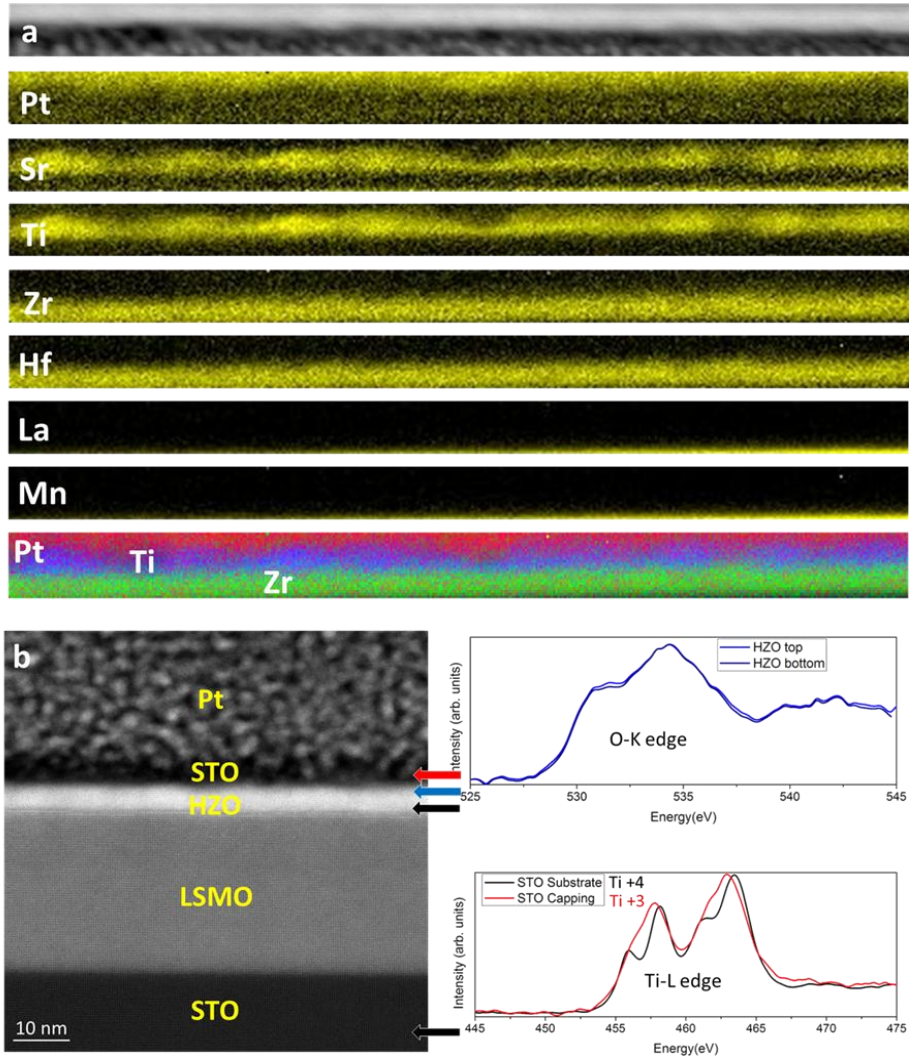

**Figure S7.** (a) Top panel shows an ADF image acquired simultaneously with EEL spectrum image from a larger area than that shown in Figure 6 in the main text. In the lower panels, the corresponding EELS elemental maps generated from the Pt-M, Sr-L3, Ti-L, Zr-L3, Hf-M, La-L and Mn-L edges. (b) Left panel, HAADF or Z-contrast image of the heterostructure. The brightest layer corresponds to the HZO layer. The arrows on the left of the image mark the region from which different spectra were obtained, see right panel. The top right panel shows the O-edge peak obtained from two different regions of the HZO layer, the uppermost region, close to the STO capping layer, and the lowermost one, just above the LSMO layer. Both peaks are similar indicating that there are no changes in the electronic structure between these two regions of the HZO layer. On the other hand, regarding the STO capping layer, we observe a reduction of the Ti when comparing its Ti-L edge with the one from the substrate (see bottom right panel). Whereas the Ti-L edge of the STO substrate has the fine structure of the $Ti^{4+}$, the Ti L-edge of the STO capping layer has the fine structure of the $Ti^{3+}$, indicating that the STO capping layer is reduced.